\documentclass[twocolumn,showpacs,preprintnumbers]{revtex4}
\usepackage{graphicx}
\usepackage{dcolumn}
\usepackage{bm}

\begin{document}

\title{Intermediate dynamics between Newton and Langevin}
\author{Jing-Dong Bao,$^{1,}$\footnote{
Electronic address: jdbao@bnu.edu.cn} Yi-Zhong Zhuo,$^2$  Fernando
A. Oliveira,$^3$ and Peter H\"{a}nggi$^{4,5}$}
\affiliation{$^1$Department of Physics, Beijing Normal University, Beijing 100875, China\\
$^2$China Institute of Atomic Energy, P. O. Box 275, Beijing 102413, China\\
$^3$Institute of Physics and International Center of Condensed
Matter
Physics, University of Brasilia-CP 04513, 70919-970, Brasilia-DF, Brazil\\
$^4$Institute of Physics, University of Augsburg, 86135 Augsburg,
Germany\\
$^5$National University of Singapore, Faculty of Science, Physics
Department, 117542 Singapore, Singapore}
\date{\today }

\begin{abstract}
A dynamics between Newton and Langevin formalisms  is elucidated
within the framework of the generalized Langevin equation. For
thermal noise yielding  a vanishing zero-frequency friction  the
corresponding non-Markovian Brownian dynamics exhibits  anomalous
behavior which is characterized by ballistic diffusion and
accelerated transport.  We also investigate the role of a possible
initial correlation between the system degrees of freedom and the
heat-bath degrees of freedom for the  asymptotic long-time behavior
of the system dynamics. As two test beds we investigate (i) the
anomalous energy relaxation of free non-Markovian Brownian motion
that is driven by a harmonic velocity noise and (ii) the phenomenon
of a net directed acceleration in noise-induced transport of an
inertial rocking Brownian motor.
\end{abstract}

\pacs{05.70.Ln, 05.40.Jc, 05.40.Ca} \maketitle

\section{introduction}

The phenomenon of Brownian motion has assumed a fundamental and
influential role in the development of  thermodynamical and
statistical theories and continues to do so as an inspiring source
for active research in various fields of natural sciences
\cite{HM2005}. The  Brownian motion dynamics can conveniently be
described by a generalized Langevin equation (GLE). The GLE was
originally derived by Mori \cite{mori1965}, Kawasaki
\cite{kawasaki73}, and Zwanzig \cite{zwanzig} by use of the
Gram-Schmidt procedure. It was further investigated by Lee using the
recurrence relations method \cite{lee2}. Starting out from the
well-known system-plus-oscillator-reservoir model as e.g. detailed
in Ref. \cite{zwanzig,HALNP}, one  obtains the GLE derived from
first principles. The validity of a thermal GLE is typically
restricted to the case with a thermal equilibrium; e.g. see in Refs.
\cite{HALNP,kubo, toda,pot}. Specifically, such a  GLE dynamics
reads \cite{kawasaki73, zwanzig,HALNP}:
\begin{equation}
m\dot{v}(t)+m\int_{0}^{t}\gamma (t-t^{\prime })v(t^{\prime
})dt^{\prime }+\partial _{x}U(x,t)=\varepsilon (t).
\end{equation}
 Notably, the thermal
noise $ \varepsilon (t)$ is  not correlated with the initial
velocity, i.e., $\langle v(0)\varepsilon(t)\rangle=0$, see in Refs.
\cite{zwanzig, HALNP} and in section III below. In contrast, the
initial position $x(0)$ typically is correlated with $ \varepsilon
(t)$. The memory friction $\gamma (t-t^{\prime })$ is in thermal
equilibrium related to the correlation of stationary random forces
\cite{kubo,toda}. Kubo \cite{kubo}
 has addressed the  common behavior of
a classical equilibrium bath by setting $\langle \varepsilon
(t)\varepsilon (t^{\prime })\rangle =mk_{B}T\gamma (t-t^{\prime})$.
Here $k_{B}$ is the Boltzmann constant and $T$ denotes the bath
temperature. The one-sided Fourier transform of $\gamma(t)$ obeys
Re$\tilde{\gamma}(\omega )\geq 0$ for  real-valued $\omega$. This
correlation result for the thermal noise $\varepsilon (t)$ is
commonly termed the fluctuation-dissipation theorem (FDT) of the
second kind \cite{toda}. The nonlinear GLE can also be extended to
account for a nonlinear system - linear bath interaction
\cite{zwanzig,HALNP, Illuminati}, yielding a structure as in Eq.
(1), but now with a nonlinear, coordinate-dependent  friction
function. It even can be generalized to arbitrary nonlinear system -
nonlinear bath interactions containing then the potential of mean
force \cite{GHT80}.

With this work we aim at  extending the theory of classical Brownian
motion by focussing  on the intricacies of a possible non-Markovian
 with an incomplete, non-Stokesian dissipative
dynamics \cite{mor,vai,lee,lutz,mok,rubi,Dhar06}. We will
demonstrate that the commonly stated  conditions for the equilibrium
bath are generally not complete within the framework of linear
response theory. This is so, because the existence of anomalous
diffusion has not been considered in the original treatment by Kubo
and others. Moreover, we discuss also the influence of initial
correlation preparation between the system and the heat bath upon
the asymptotical behavior of the force-free system.

\section{Biasing generalized Brownian motion}

Let us first consider a  free Brownian dynamics with $U(x,t)=0$,
possessing via the FDT of the second kind a finite-valued
zero-frequency friction. If this dynamics is next subjected to a
constant external force, i.e. $-\partial U(x,t)/\partial x = F $,
the acceleration vanishes in the case of a Stokesian friction,
because the external force balances the friction force. A problem of
broad interest is whether there exists an \textit{intermediate
situation} between the Newtonian mechanics and such an ordinary
Langevin formalism. This in turn necessitates a non-Stokesian
dissipation mechanism such that the asymptotic long-time statistical
probability will typically approach a stationary state that
explicitly depends on the initial preparation. It is thus of great
practical interest to research what kind of heat bath can take on
this role. Such non-ergodic non-equilibrium thermodynamics presents
a timely subject that is presently hotly debated, both within theory
\cite{mor,vai,lee,lutz,Muk2005,lee2006,Bai2005} and experiment
\cite{Kutnjak99,Fahri99,Brok03}.

For a GLE subjected to a constant force; i.e.  $U(x,t)= - Fx$, the
solution of (1) can be written as
\begin{eqnarray}
x(t)&=&x(0)+v(0)H(t)+\frac{1}{m}\int^{t}_0dt'H(t-t')(\varepsilon
(t')+F),\nonumber\\
v(t)&=&v(0)h(t)+\frac{1}{m}\int^{t}_0dt'h(t-t')(\varepsilon (t')+F),
\end{eqnarray}
where $H(0)=0$ and $h(0)=1$. The two response functions $H(t)$ and
$h(t)$ are the inverse of the Laplace transforms
$\hat{H}(s)=[s^2+s\hat{\gamma}(s)]^{-1}$ and
$\hat{h}(s)=[s+\hat{\gamma}(s)]^{-1}$, respectively, where
$\hat{\gamma}(s)$ is the Laplace transform of memory friction
kernel, i.e. $\hat{\gamma}(s)=\int^{\infty}_0\gamma(t)\exp(-st)dt$.
Under the  assumption  that the characteristic equation
$s+\hat{\gamma}(s)=0$ possesses a zero root, i.e. $s=0$,  the
residue theorem implies that
$h(t)=b+\sum_{i}\textmd{res}[\hat{h}(s_i)]\exp(s_it)$ and $H(t)=c+bt
+\sum_is^{-1}_i\textmd{res}[\hat{h}(s_i)]\exp(s_it)$. Here, $s_i$
(Re$s_i<0$) denote  the non-zero roots of the above characteristic
equation and res[$\cdots$] are the residues. Within this context,
the two relevant, generally non-vanishing quantities $b$ and $c$ are
determined to read:
\begin{equation}
b=\frac{1}{1+\hat{\gamma}'(0)},\quad
c=-\frac{1}{2}b^2\hat{\gamma}''(0).
\end{equation}
 This result then requires that
$\hat{\gamma}(0)=\int^{\infty}_0\gamma(t)dt=0$, i.e.,
 implying a vanishing
effective friction at zero frequency.

 The average velocity
and the average displacement under the  external bias $F$ emerge at
long times as
\begin{eqnarray}
\{\langle v(t\to\infty)\rangle\}&=&b\left(\{
v(0)\}+\frac{F}{m}t\right)+\frac{F}{m}c,\\
\{\langle
x(t\to\infty)\rangle\}&=&\{x(0)\}+b\left(\{v(0)\}t+\frac{1}{2}\frac{F}{m}t^2\right)\nonumber\\
&+&c\left(\{v(0)\}+\frac{F}{m}t\right) +\frac{F}{m}d,
\end{eqnarray}
where $d=-\sum_{i}\textmd{res}[\hat{h}(s_i)]/s^2_i$ is a
noise-dependent quantity. Herein, we indicate by $\{\cdots \}$ the
average with respect to the initial preparation of the state
variables, i.e. an average over their initial values and $\langle
\cdots\rangle$ is the noise average. The dissipative {\it
acceleration} of a Brownian particle of mass $m$ subjected to a
constant force $F$ then reads:
\begin{equation}
a=\frac{F}{m}b,
\end{equation}
 where generally  $0\leq b\leq 1$. This quantity $b$ will be termed the
 \textit{dissipation reducing} factor, the dissipation is
reduced as $b$ is increased.
 This result is  intermediate
between a purely  Newtonian mechanics (obeying $b=1$) and an
ordinary Langevin dynamics (with $b=0$) including of  GLEs.

 The two limiting results for the asymptotic
dynamics are found to read: (i) The Newton case with $b=1$ implying
no dissipation, i.e. $\gamma(t)\equiv 0$ with $c=d=0$, yielding:
$a=F/m$, $v(t)=v(0)+\frac{F}{m}t$, and
$x(t)=x(0)+v(0)t+\frac{1}{2}\frac{F}{m}t^2$. (ii) The commonly
known, ordinary Langevin situation is obtained with  $b=0$; i.e.
Eq. (3) then looses validity because of $\hat{\gamma}(0)\neq 0$. For
this case $c=\hat{\gamma}^{-1}(0)$ where $\hat{\gamma}(0)$ denotes
the Markovian friction strength, resulting in  $a=0$, $\langle
v(t\to\infty)\rangle=F/(m\hat{\gamma}(0))$, and $\{\langle
x(t\to\infty)\rangle\}=\{x(0)\}+\{v(0)\}/\hat{\gamma}(0)+\frac{F}{m}t/\hat{\gamma}(0)+\frac{F}{m}d$.

In the unbiased case, we derive the two-time velocity correlation
function (VCF) of free generalized Brownian motion in a generic
form, i.e.,
\begin{eqnarray}
\{\langle
v(t_{1})v(t_{2})\rangle\}&=&\frac{k_BT}{m}h(|t_1-t_2|)+\left(\{v^2(0)\}-\frac{k_BT}{m}\right)\nonumber\\
&&h(t_1)h(t_2).
\end{eqnarray}
Here we used only the condition: $\langle
v(0)\varepsilon(t)\rangle=0$.  Note that depending on the specific
choice for the initial preparation this velocity  correlation
generally is not time-homogeneous. The stationary velocity
correlation function becomes again only a function
 of $|t_1-t_2|$  for the case that we use the equilibrium
 preparation with an
initial velocity variance in accordance with the thermal equilibrium
value, i.e.   $\{v^2(0)\}=k_{B}T/m$ \cite{Bai2005}. The deduced
asymptotic stationary VCF then reads:
$C_{vv}(\infty)=b\frac{k_BT}{m}\neq 0$. This causes a breakdown of
the ergodic equilibrium state because of the initial
preparation-dependence, which is encoded in the $v(0)$-dependent
asymptotic results: $\langle v\rangle_{st}=bv(0)\neq 0$ and
$\{\langle v^2\rangle\}_{st}=k_{B}T/m+b^2(\{v^2(0)\}-k_BT/m)$.

Likewise, the mean square displacement (MSD) of the force-free
particle is written as
\begin{eqnarray}
\{\langle
x^2(t)\rangle\}&=&\{x^2(0)\}+\{v^2(0)\}H^2(t)+2\{x(0)v(0)\}H(t)\nonumber\\
&+&\frac{2}{m}\int^{t}_0dt'H(t-t')\{\langle
x(0)\varepsilon(t')\rangle\}\nonumber\\
&+&\frac{k_BT}{m}\left(2\int^t_0H(t')dt'-H^2(t)\right),
\end{eqnarray}
where the fourth term denotes the effect of initial coupling between
system and heat bath. We will discuss the correlation preparation in
the following  section. Note that here the largest power in the
temporal variation of the MSD involves the square of time. The
averaged displacement can be related to the MSD via the generalized
Einstein relation, reading
\begin{equation}
\kappa_2/ \mu_2(F\rightarrow 0)=k_BT_{\textmd{eff}} \;.
\end{equation}
The {\it ballistic} diffusion coefficient is $
\kappa_2=\lim_{t\to\infty}\{\langle x^2(t)\rangle\}_{F=0}/(2t^2)$,
being related to the increasing rate of linear mobility
$\mu_2(F\rightarrow 0)=\lim_{t\to\infty}\{\langle x(t)\rangle\}/(F
t^2)_{F=0}$. Here, this  effective temperature formally reads:
$T_{\textmd{eff}}:=T+b(m\{v^2(0)\}/k_B - T)$, where $T$ is the
temperature for the common case with $b=0$.

To assure the equilibrium behavior of this generalized Brownian
motion  the usual condition of Kubo's FDT of the second kind for the
thermal noise must be complemented as follows: Consider the Fourier
transform $\tilde{\gamma}(\omega)$ of the memory damping kernel,
i.e.,
\begin{equation}
\tilde{\gamma}(\omega)=\hat{\gamma}(s=-i\omega),
\end{equation}
The real part of the former quantity is the spectral density of
noise. A ballistic diffusion with $b\neq 0$ thus requires that the
lowest power of $\hat{\gamma}(s)$ is of first-order in $s$, implying
that the lowest power of Re$\tilde{\gamma}(\omega)$ is proportional
to $\omega^2$ at low frequencies. Therefore, for a genuine thermal
noise driven, force-free particle approaching at the equilibrium
state the usual conditions must be  completed by: $\lim_{s\to
0}(\hat{\gamma}(s)/s)\to\infty$, or $\lim_{\omega\to
0}(\textmd{Re}\tilde{\gamma}(\omega)/\omega^2)\to\infty$.

\section{INITIAL CORRELATION BETWEEN SYSTEM AND BATH}

Starting from the system-plus-reservoir model, one knows that the
coupling between the system and environmental degrees of freedom
there exist four kinds of coupling forms which do not involve a
renormalization of potential or mass of the system. In these cases
the heat bath consists of  a set of independent harmonic oscillators
with  masses $m_j$ and oscillation frequencies $\omega_j$. Pervious
work \cite{Bai2005} has shown that the random force is {\it
independent} of the system variables for the coordinate-velocity
coupling. Nevertheless, however, the expression of thermal noise
will depend on the initial preparation of system for the coupling
between the system coordinate (velocity) and the environmental
coordinates (velocities) considered here. To the best of our
knowledge, only a small number of prior studies \cite{gra,HT} have
considered the general consequences of the detailed initial
preparation procedure in view of the asymptotic statistical results
of the system.

\subsection{The coordinate-coordinate coupling}

For a bilinear coupling between the system coordinate $x$ and the
heat bath's coordinates $q_j$, the total Hamiltonian can be written
as:
\begin{equation}
H=\frac{P_x^2}{2m}+U(x,t)+\sum_j\left[\frac{p^2_j}{2m_j}+\frac{1}{2}m_j\omega^2_j\left(
q_j-\frac{c_j}{m_j\omega^2_j}x\right)^2\right]\;.
\end{equation}
Here and below the momenta of the system and bath's oscillators are
related to $P_x=mv$ and $p_j=m_j\dot{q}_j$, respectively, and the
set $c_j$ denote the coupling constants.
 The equation of motion of the system obeys the form of the GLE (1) and
 the thermal noise  appearing in Eq. (1) emerges
as \cite{zwanzig,HALNP,Bai2005}
\begin{equation}
\varepsilon(t)=-m\gamma(t)x(0)+\xi_{\textmd{bath}}(t),
\end{equation}
where
$\gamma(t)=\frac{1}{m}\sum_j\frac{c^2_j}{m_j\omega^2_j}\cos\omega_jt$
 and $\xi_{\textmd{bath}}(t)$ is
determined by the initial coordinates and the velocities of the
oscillators of the heat bath. The bath part $\xi_{\textmd{bath}}(t)$
of the noise explicitly reads:
\begin{equation}
\xi_{\textmd{bath}}(t)=\sum_jc_j\left(q_j(0)\cos\omega_jt+\frac{\dot{q}_j(0)}{\omega_j}\sin\omega_jt\right).
\end{equation}

Physically, this thermal noise $\varepsilon(t)$ obeys statistical
properties that  derive from the canonical,  thermal equilibrium
distribution of the total, combined system-plus-bath
\cite{zwanzig,HALNP, ros}: this thermal noise then again yields  a
vanishing mean and its correlation obeys the thermal FDT
\cite{HALNP}. The statistical quantities involving noise the system
variables are strictly determined by the joint probability
\cite{gra}. The correlation involving initial position of the system
and the thermal noise $\varepsilon (t')$, i.e., the fourth term in
Eq. (8) reads
\begin{eqnarray}
&&\frac{2}{m}\int^{t}_0dt'H(t-t')\{\langle
x(0)\varepsilon(t')\rangle\}\nonumber\\&=&
-2\{x^2(0)\}\int^t_0dt'H(t-t')\gamma(t')\nonumber\\
&=&-2H(t)\ast\gamma(t)\{x^2(0)\}=-2(1-h(t))\{x^2(0)\},\nonumber\\
\end{eqnarray}
 wherein
$``\ast"$ denotes the convolution integral. It reads,
$H(t)\ast\gamma(t)=\frac{1}{2\pi i}\int
ds\hat{H}(s)\hat{\gamma}(s)\exp(st)=1-\dot{H}(t)=1-h(t)$.  Here we
have used the relation below Eq. (2), i.e.,
$\hat{H}(s)\hat{\gamma}(s)=s^{-1}-s\hat{H}(s)$. This  contribution
assumes a finite value, i.e.,  $-2(1-b)\{x^2(0)\}$ in the long-time
limit.

Under the usual assumption  that the thermal noise $ \varepsilon
(t)$ and the initial velocity $v(0)$ of the system are not
correlated, we find from Eq. (12) that the initial coordinate of the
system must be uncorrelated with its initial velocity for the
coordinate-coordinate coupling case, namely, $\{x(0)v(0)\}=0$, being
the case for a canonical thermal equilibrium, cf. the Hamiltonian in
Eq. (11). Therefore,  the third term in Eq. (8) also vanishes. In
the following we shall not consider preparations with  such initial
correlations between the initial coordinate $x(0)$ and the initial
velocity  $v(0)$.

\subsection{The velocity-velocity coupling}

For a bi-linear coupling between the system velocity and the
velocities of the bath oscillators the total Hamiltonian reads
\begin{equation}
H=\frac{P_x^2}{2m}+U(x,t)+\sum_j\left[\frac{1}{2m_j}\left(p_j-\frac{d_j}{m}P_x\right)^2+\frac{1}{2}m_j\omega^2_j
q_j^2\right],
\end{equation}
where $d_j$ denote the corresponding  the coupling constants. We can
derive again the GLE (1) describing the motion of the system with
the thermal noise term now given by
\begin{eqnarray}
\varepsilon(t)
&=&v(0)\sum_j\frac{d^2_j\omega_j}{m_j}\sin\omega_jt\nonumber\\
&+&\sum_j
d_j\omega^2_j\left[q_j(0)\cos\omega_jt+\frac{\dot{q}_j(0)}{\omega_j}\sin\omega_jt\right]\nonumber\\
&=& v(0)m\int^t_0\gamma(t')dt'+\xi_{\textmd{bath}}(t),
\end{eqnarray}
where
$\gamma(t)=\frac{1}{m}\sum_j\frac{d^2_j\omega^2_j}{m_j}\cos\omega_jt$
 and in addition  we have: $\{\langle
x(0)\xi_{\textmd{bath}}(t)\rangle\}=\{\langle
v(0)\xi_{\textmd{bath}}(t)\rangle\}=0$. We require that the FDT of
the second kind is obeyed, namely that  $\langle
\varepsilon(t)\rangle=0$ and $\langle
\varepsilon(t)\varepsilon(t')\rangle=k_BT\gamma(t-t')$. This is
guaranteed when $\langle q_j(0)\rangle=\langle
\tilde{\dot{q}}_j(0)\rangle=\langle
q_i(0)\tilde{\dot{q}}_j(0)\rangle=0$, $\langle
q_i(0)q_j(0)\rangle=\frac{k_BT}{m\omega^2_j}\delta_{ij}$, and
$\langle
\tilde{\dot{q}}_i(0)\tilde{\dot{q}}_j(0)\rangle=\frac{k_BT}{m_j}\delta_{ij}$,
where $\tilde{\dot{q}}_i(0)=\dot{q}_i(0)-d_j/m_j v(0)$
\cite{Bai2005}.

In the case of velocity-velocity coupling, the fourth term in Eq.
(8) vanishes if again $\{x(0)v(0)\}=0$. An additional term emerges,
however, for the mean squared displacement (MSD) of the force-free
particle due to the thermal noise $\varepsilon(t)$ which now depends
on the initial particle velocity $v(0)$.  From Eq. (2), we obtain
\begin{eqnarray}
\{\langle
x^2(t)\rangle\}_{\textmd{add}}&=&\frac{2}{m}H(t)\int^t_0dt'H(t-t')\{\langle
v(0)\varepsilon(t')\rangle\}\nonumber\\
&=&2H(t)\{v^2(0)\}\int^t_0dt'H(t-t')\int^{t'}_0du\gamma(u)\nonumber\\
&=&2(t-H(t))H(t)\{v^2(0)\}.
\end{eqnarray}
Indeed, the ballistic diffusion arises also in this case. Notably,
an additional contribution to the mean square velocity of the
force-free particle [c.f. Eq. (7) at $t_1=t_2$] emerges in this
case. It reads
\begin{eqnarray}
\{\langle
v^2(t)\rangle\}_{\textmd{add}}&=&\frac{2}{m}h(t)\int^t_0dt'h(t-t')\{\langle
v(0)\varepsilon(t')\rangle\}\nonumber\\
 &=&2\{v^2(0)\}h(t)\int^t_0dt'h(t-t')\int^{t'}_0\gamma(u)du\nonumber\\
 &=&2h(t)(1-h(t))\{v^2(0)\},
\end{eqnarray}
where we have used the inverse Laplace transform of the convolution
integral in Eq. (18) by making use of the relation,
$\hat{h}(s)\hat{\gamma}(s)=1-s\hat{h}(s)$.

In particular, the mean squared velocity of the force-free particle
emerges in the  long-time limit as
\begin{eqnarray}
\{\langle
v^2(t\to\infty)\rangle\}&=&2b(1-b)\{v^2(0)\}+\frac{k_BT}{m}\nonumber\\
&+&\left(\{v^2(0)\}-\frac{k_BT}{m}\right)b^2.
\end{eqnarray}
This result evidences that the system can not arrive at the
equilibrium state for any initial preparation of the particle
velocity if $b\neq 0$. Therefore, for the validity of FDT one has to
use at initial time a preparation of thermal equilibrium for the
system and the heat bath. Nevertheless, one needs not to worry that
the noise is uncorrelated with the initial velocity of the system
for a common non-Markovian dynamics with $b=0$. Then, the FDT is
valid  independent of the coupling form between system and bath
whenever the effective Markovian damping of the system is finite at
zero frequency. This is so because the first and the third term in
Eq. (19) vanishes for $b=0$.

\section{Test bed for non-Stokesian  dissipative dynamics}

Given the FDT of the second kind by Kubo  we investigate next
unbiased, non-Markovian Brownian motion in (1) that is driven by
colored noise known as the harmonic velocity noise (HVN)
$\varepsilon(t)$ \cite{bao2005}, which, however,  does not obey the
above additional requirements. The HVN itself is produced  from a
linear Langevin equation, namely,
\begin{equation}
\dot{y}=\varepsilon, \quad \dot{\varepsilon}=-\Gamma
\varepsilon-\Omega ^{2}y+\xi(t),
\end{equation}
 where $\xi(t)$ denotes  Gaussian white noise of vanishing
mean with $\langle \xi (t)\xi (t^{\prime })\rangle =2\eta\Gamma
^{2}k_{B}T\delta (t-t^{\prime})$. The coefficient $\eta $ denotes
the damping coefficient of the system corresponding to the thermal
white noise; $\Gamma$ and $\Omega $ denote the damping and the
frequency parameters. The second moments and the cross-variance of
$y(0)$ and $\varepsilon(0)$ obey: $ \{y^{2}(0)\}=\eta \Gamma \Omega
^{-2}k_{B}T$, $\{\varepsilon^{2}(0)\}=\eta \Gamma k_{B}T$, and
$\{y(0)\varepsilon(0)\}=0$. The Laplace transformation of the memory
damping kernel reads $\hat{\gamma}(s)=\eta\Gamma s/(s^2+\Gamma
s+\Omega^2)$ with
\begin{equation}
\textmd{Re}\tilde{\gamma}(\omega)=\frac{\eta\Gamma^2\omega
^{2}}{(\Omega ^{2}-\omega ^{2})^{2}+\Gamma ^{2}\omega ^{2}}\;,
\end{equation}
respectively. The latter corresponds to the spectrum of HVN which
indeed vanishes identically at zero-frequency. In this case the
dissipation reducing factor emerges as  $b=(1+\eta \Gamma \Omega
^{-2})^{-1}$, and likewise, $c=(1-b)^2/\eta$,
$d=(1-b)^2[(\eta\Gamma)^{-1}-(1-b)\eta^{-2}]$.

Using models with a bi-linear coordinate system-bath coupling the
dynamics can be characterized  by the  spectral density of bath
modes, $J(\omega)$, being related to
Re$\tilde{\gamma}(\omega)=J(\omega)/ m\omega$
\cite{HTB,gra87,RI,chen}. Thus, for a weak coupling to a bath, as it
can be realized either with optical-like bath modes \cite{mor, RI},
broadband colored noise \cite{bao2003}, or also for the  celebrated
case of a black-body radiation field of the Rayleigh-Jeans type
\cite{for} the static friction  vanishes. Yet other physical
situations that come to mind involve the vortex diffusion in
magnetic fields \cite{ao}, or  open system dynamics with a
velocity-dependent system-bath coupling
\cite{for,Bai2005,bao06,poll}.

The non-Markovian Brownian motion can equivalently  be recast as an
embedded, higher-dimensional Markovian process. The Fokker-Planck
equation (FPE) for the  probability density
$P(x,v,w,u,y,\varepsilon;t)$ corresponds then to the dynamics of a
set of coupled Markovian LEs involving the auxiliary-variables
$(w,u,y,\varepsilon)$: It obeys $\partial_{t}P = L_{FP}P$, where
$L_{FP}$ is the associated
 FPE operator; when being formally supplemented here with a nonvanishing
 potential $U(x,t)$, it explicitly reads:
\begin{eqnarray}
L_{FP}&=& -v\frac{\partial}{\partial
x}-m^{-1}(-U'(x,t)+w)\frac{\partial}{\partial v}\nonumber\\
&&+(\Gamma w+\eta\Gamma v+\Omega^2y+u)\frac{\partial}{\partial
w}\nonumber\\
&&-\Omega^2(w-\varepsilon)\frac{\partial}{\partial
u}+\varepsilon\frac{\partial}{\partial y}\nonumber\\
&&+(\Gamma \varepsilon+\Omega^2 y)\frac{\partial}{\partial
\varepsilon}+\eta\Gamma^2 k_BT\frac{\partial^2}{\partial
w^2}+\eta\Gamma^2 k_BT\frac{\partial^2}{\partial
\varepsilon^2}.\nonumber\\
\end{eqnarray}

\begin{figure}[tbp]
\includegraphics[width=9cm,height=8cm]{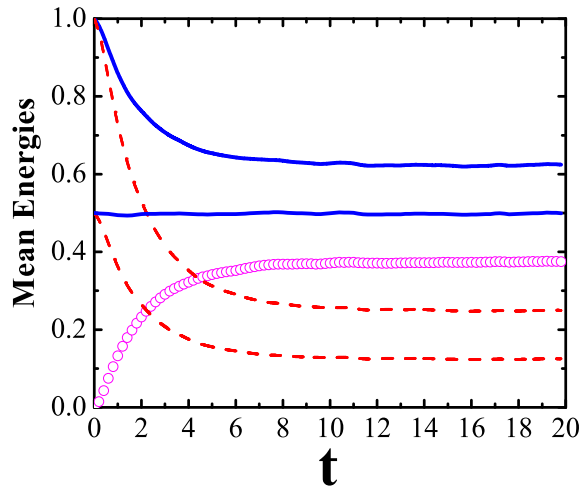}
\caption{(color online). Behavior of diverse mean energies (see
text) for a nonergodic, force-free Brownian particle {\it vs.} time
$t$. The parameters used are $m=1.0$, $k_BT=1.0$, $\eta=0.2$,
$\Gamma=5.0$, $\Omega=1.0$, $x(0)=0$, and (i) $v(0)=\sqrt{2}$ and
(ii) $1.0$, respectively, for the {\it total} energies given by the
solid lines and the {\it remnant} energies by the dashed lines, from
top to bottom. The open circles is the adsorbed energy from the bath
for all cases and also the total energy for the $v(0)=0$ case.}
\label{1}
\end{figure}

\subsection{The energy relaxation}

We use this form for the analysis of the energy relaxation. The mean
total energy of the thermal HVN-driven force-free particle reads
\begin{eqnarray}
\{\langle E(t)\rangle\}&=&\frac{1}{2}m\{\langle
v^2(t)\rangle\}\nonumber\\
&=&\frac{1}{2}m\{\langle
v(t)\rangle^2\}+\frac{1}{2}m\{\langle(v(t)-\langle
v(t)\rangle)^2\rangle\}.\nonumber\\
\end{eqnarray}
 The first part describes the remnant
initial kinetic energy of the particle, being dissipated partly by
the  heat bath environment. This part  vanishes in the ordinary case
with $b=0$. The second part denotes the energy provided from the
heat bath. It is independent of the initial particle velocity, but
does not relax, however, towards equilibrium. The absorbed power of
the particle from the heat bath, namely, the rate of work being done
by the fluctuation force \cite{li}, is
\begin{eqnarray}
P_{\textmd{abs}}&=&\lim_{t\to\infty}\int^t_0\gamma(t-t')\{\langle
v(t)v(t')\rangle\}dt'\nonumber\\
 &=&\frac{\eta
k_{B}T}{1+\eta/(2\Gamma)+2\Omega^2/(\eta\Gamma)} \;.
\end{eqnarray}
 Note that it falls short of the equilibrium value $\eta k_{B}T$.
All quantities plotted are dimensionless.
 Our numerical results
are depicted with Fig. 1.  These results are obtained via the
simulation of a set of Markovian LE's which are equivalent to the
FPE in (22) with $U(x,t)=0$.

\begin{figure}[tbp]
\includegraphics[width=9cm,height=8cm]{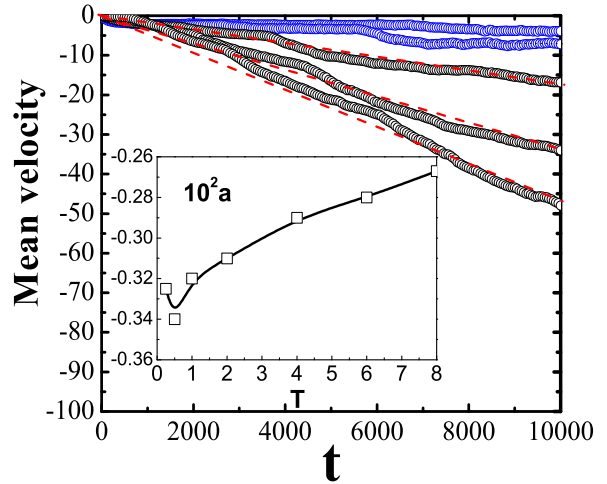}
\caption{(color online). Accelerated, (time and ensemble)- averaged
Brownian motor velocity  in a rocking ratchet  that is driven by
HVN. The used parameters are $x(0)=v(0)=0$, $m=1.0$, $k_BT=0.5$,
$\protect\eta =5.0$, $\Gamma=22.0$, $\Omega^2=40.0$, $t_p=25.0$,
$A_0=2.0$, $4.0$, $6.0$, $10.0$, and $15.0$, from top to bottom. The
particles undergo a finite acceleration  $a$ (times 100) {\it vs.}
temperature $T$, being depicted with  the inset  for $A_0=10.0$.}
\label{1}
\end{figure}

\subsection{Brownian motor exhibiting accelerated transport}

A most intriguing situation refers to Brownian motors \cite{BM} when
driven by Brownian motion that exhibits ballistic diffusion. Take
the case of thermal HVN driving a Brownian particle according to (1)
with a periodic potential that breaks reflection symmetry, namely
\begin{equation}
U(x,t)=U_0[\sin(2\pi x)+c_1\sin(4\pi x)+c_2\sin(6\pi x)]+A(t)x,
\end{equation}
 where $U_0=0.461$, $c_1=0.245$, and $c_2=0.04$
\cite{mac}. $A(t)$ is a square-wave periodic driving force that
switches forth and back among $A(t)=A_{0}$ when $2nt_{p}\leq
t<(2n+1)t_{p}$ and $A(t)=-A_{0}$ when $(2n+1)t_{p}\leq
t<2(n+1)t_{p}$. The key challenge is whether a non-vanishing
non-equilibrium current emerges that can be put to a constructive
use in order to direct, separate or shuttle particles efficiently
\cite{BM}.

Figure 2 depicts the  ensemble- and driving-phase-averaged velocity
(with the latter being equivalent to an average over the temporal
driving period, see in Ref. \cite{JH91}), i.e.,
\begin{equation}
\overline{\{\langle v(t)\rangle\}}=(2t_{p})^{-1}\int_{t}^{t+2t_{p}}
dt^{\prime}\{\langle v(t^{\prime })\rangle\}
\end{equation}
 for various strengths
of the driving force $A_0$, as obtained via the simulation of the
Markovian LE's corresponding to the equivalent higher-dimensional
FPE in (22). A startling finding is now that this averaged velocity
is no longer a constant but rather increases {\it linearly} with
time. This is in clear contrast to the behavior of ordinary Brownian
motors \cite{BM}. This {\it directed acceleration} is presented by
the slope (see dashed lines in Fig. 2) of the average velocity. For
a weak rocking (i. e. small $A_{0}$) the phenomenon of directed
motion involves the surmounting of barriers, thus hindering
transport. In contrast, for a strong super-threshold rocking the
averaged displacement is related to the mean square displacement via
the modified Einstein relation in  (9), yielding $\{\langle
x(t)\rangle\}\propto - F_{\textmd{eff}} \; t^2$ in the long-time
limit. Here $F_{\textmd{eff}}$ is an effective tilting force
stemming from the rocking of the ratchet potential. Amazingly, this
Brownian motor can be accelerated because the driving and the
noise-induced effective tilting force supersedes the acting friction
force.

\section{Conclusions}

We have researched within the GLE-formalism in (1) an intermediate
dynamics proceeding between Newton and Langevin. The emerging
non-equilibrium features are manifested by the initial
preparation-dependent asymptotic stationary state, which is directly
related to a non-Stokesian dissipative phenomenon which stems from a
vanishing effective Markovian friction at zero-frequency. It has
been found that the fluctuation-dissipation theory is valid and
independent of the coupling form between system and bath when the
effective Markovian damping is finite at zero frequency.
 In order to assure the equilibrium behavior of a system,
 the usual condition for the heat bath, i.e., the Kubo's fluctuation-dissipation
theorem of the second kind,  must be completed by an additional
requirement: $\lim_{s\to 0}(\hat{\gamma}(s)/s)\to\infty$ or
$\lim_{\omega\to
0}(\textmd{Re}\tilde{\gamma}(\omega)/\omega^2)\to\infty$, where
$\hat{\gamma}(s)$ and $\tilde{\gamma}(\omega)$ are the Laplace and
Fourier transforms of the memory damping. However, the system can
not arrive at the equilibrium state for any initial preparation of
the system if the the condition is not obeyed for the bilinear
coupling between the system velocity and the velocities of
environmental oscillators.

Our findings exhibit anomalous super-diffusion in the form of a
ballistic diffusion. Yet another riveting result is that the
corresponding Brownian dynamics for a rocking Brownian motor
exhibits a distinct, {\it accelerated} velocity, rather than the
constant drift which typifies the situation with a Stokesian finite
zero-frequency dissipation. We are also confident that our present
results will serviceably impact other quantities of thermodynamic
and quantum origin. Thus, this field is open for future studies that
in turn may reveal further surprising findings.

\section*{ACKNOWLEDGEMENTS}

This work was supported by the NNSFC under 10674016 and 10475008 and
the German Research Foundation (DFG), Sachbeihilfe HA1517/26-1 (P.
H.) and the CNPq (Brazil) (F. A. O.).

\end{document}